\documentclass[acmsmall]{acmart}
\AtBeginDocument{%
  }

\setcopyright{acmlicensed}
\copyrightyear{2025}
\acmYear{2025}
\acmDOI{XXXXXXX.XXXXXXX}

\acmJournal{JACM}
\acmVolume{37}
\acmNumber{4}
\acmArticle{111}
\acmMonth{8}

\usepackage[ruled,linesnumbered]{algorithm2e}
\usepackage{algorithmic}
\usepackage{amsmath}
\usepackage{subcaption} 
\usepackage{graphicx}
\usepackage{multirow}
\usepackage{bm}

\begin{document}

\title{Real-Time Network Traffic Forecasting with Missing Data: A Generative Model Approach}

\author{Lei Deng}
\email{leideng@hkust-gz.edu.cn}
\author{Wenhan XU}
\email{wxube@connect.ust.hk}
\author{Jingwei Li}
\email{jingweili@hkust-gz.edu.cn}
\author{Danny H.K. Tsang}
\authornotemark[1]
\email{eetsang@ust.hk}
\affiliation{%
  \institution{The Hong Kong University of Science and Technology (Guangzhou)}
  \city{Guangzhou}
  \state{Guangdong}
  \country{China}
}

\renewcommand{\shortauthors}{Deng et al.}

\begin{abstract}
Real-time network traffic forecasting is crucial for network management and early resource allocation. Existing network traffic forecasting approaches operate under the assumption that the network traffic data is fully observed. However, in practical scenarios, the collected data are often incomplete due to various human and natural factors. In this paper, we propose a generative model approach for real-time network traffic forecasting with missing data. Firstly, we model the network traffic forecasting task as a tensor completion problem. Secondly, we incorporate a pre-trained generative model to achieve the low-rank structure commonly associated with tensor completion. The generative model effectively captures the intrinsic low-rank structure of network traffic data during pre-training and enables the mapping from a compact latent representation to the tensor space. Thirdly, rather than directly optimizing the high-dimensional tensor, we optimize its latent representation, which simplifies the optimization process and enables real-time forecasting. We also establish a theoretical recovery guarantee that quantifies the error bound of the proposed approach. Experiments on real-world datasets demonstrate that our approach achieves accurate network traffic forecasting within 100 ms, with a mean absolute error (MAE) below 0.002, as validated on the Abilene dataset.
\end{abstract}



\maketitle

\section{Introduction}
The widespread adoption of smart devices has significantly accelerated the global expansion of the Internet, leading to a sharp increase in both network traffic and application demands. This rapid growth has amplified the complexity of the network and the volume of data that require continuous monitoring and management \cite{xie2023deep,deng2023}. In the progress toward intelligent networks, one of the crucial challenges is real-time network traffic forecasting. This task involves estimating future traffic volumes based on historical data to prevent congestion proactively and maintain high network performance \cite{aouedi2025deep}. Real-time forecasting plays a vital role in enabling the network operator to promptly assess user demands, identify anomalous or malicious activities, and implement dynamic resource allocation strategies.

Network traffic forecasting heavily depends on historical data, which provides essential spatiotemporal information for accurate predictions. The quality of this historical data significantly influences the forecasting performance. Unfortunately, in real-world scenarios, such data are often incomplete due to various human and natural factors: (i) to reduce measurement costs, normally only a subset of origin-destination (OD) pairs would be selected for taking traffic measurements \cite{zhu2016network,deng2023}, resulting in partial observations; (ii) due to system-related issues \cite{xie2018accurate} such as network congestion, node misbehavior, transmission interference, and monitor failure can lead to inevitable data loss \cite{xie2018accurate,tan2021packet,tan2021band}. Therefore,
achieving real-time and accurate network traffic forecasting in the presence of missing data has become a pressing challenge.

Existing research on network traffic forecasting mainly includes two categories: traditional statistical techniques and deep learning (DL)-based approaches. The majority of statistical models rely on linear methods, such as Autoregressive (AR) and Auto-Regressive Integrated Moving Average (ARIMA) models \cite{moayedi2008arima}. While ARIMA is capable of modeling linear and short-range dependencies (SRD), it fails to capture long-range dependencies (LRD), often leading to poor performance in network traffic forecasting \cite{gao2021accurate}. DL methods typically incorporate advanced architectures, including graph neural networks and Long Short-Term Memory (LSTM) models, to effectively learn spatial-temporal correlations within network traffic data, enabling more precise forecasting \cite{xie2024m,Nihale2020lstm,saha2023analyzing,pandey20245gt}. However, most of these methods assume the availability of complete historical data, which is rarely the case in real-world deployments. As a result, the above approaches cannot be directly applied for network traffic forecasting with missing data. Although this problem can be addressed by applying data imputation techniques prior to forecasting, this two-step process can distort key latent structures in the data and may result in cumulative forecasting errors. It is desirable to develop a new approach for real-time network traffic forecasting with missing data.

Real-time network traffic forecasting becomes challenging in the presence of missing data.  Specifically, network traffic data often exhibits a low-rank structure, reflecting its underlying spatiotemporal correlations \cite{xie2023deep,deng2023}. However, missing entries can severely distort this inherent structure \cite{xie2018accurate,deng2021}, making it difficult to effectively leverage low-rank properties during forecasting. This disruption not only complicates the exploitation of spatial-temporal patterns but also slows down the optimization process, thereby hindering the ability to perform real-time forecasting.


With the rapid advancement of generative artificial intelligence \cite{goodfellow2014generative,liu2021real,yang2023adv}, generative models have emerged as potential tools for enforcing structural constraints in network traffic data. Notably, pre-trained generative models have demonstrated strong capabilities in recovering sparse signals from limited measurements in the context of compressive sensing \cite{bora2017compressed,wu2019deep,jalal2020robust}. Inspired by these successes, we consider a pre-trained generative model to capture the intrinsic low-rank structure of network traffic data, which can generate a complete traffic tensor from its latent representation. With the help of generative models, it becomes feasible to effectively exploit the underlying low-rank structure of traffic data even in the presence of missing entries, thereby facilitating accurate and real-time traffic forecasting.



In this paper, we propose a Generative Model approach for real-time network traffic forecasting with missing data (GMF). Firstly, we model the task of network traffic forecasting with missing data as a tensor completion (TC) problem \cite{zhang2016exact,deng2020}. Secondly, we propose a GMF scheme for real-time network traffic forecasting, which employs a pre-trained generative model to relax the low-rank constraint typically associated with traffic tensor completion. The pre-trained generative model allows GMF to learn the low-rank patterns of traffic tensors in the pre-training stage and can map a condensed latent representation to a complete traffic tensor. GMF circumvents the requirement of low-rank tensor decomposition and simplifies the optimization process in TC, leading to fast forecasting.

Our main contributions are summarized as follows:
\begin{itemize}
\item
We propose a general TC framework based on generative models. This framework leverages a generative model to impose low-rank structures that are fundamental to TC, which maps a low-dimensional latent representation to a low-tubal-rank tensor. This design decouples the low-rank constraint from the tensor optimization, allowing real-time TC by updating only the compact latent representation.

\item
We propose a Generative Model approach for real-time network traffic Forecasting with missing data (GMF). GMF reformulates the forecasting task as a low-rank TC problem and enables real-time forecasting by mapping a compact latent representation to a complete low-rank traffic tensor. The neural architecture of the generative model in GMF is carefully designed, incorporating novel tensor layers that effectively capture low-rank structures, thereby enhancing the accuracy and reliability of traffic forecasting.

\item
We perform extensive evaluations on real-world network traffic datasets to validate the proposed GMF scheme. The experimental outcomes consistently demonstrate that GMF surpasses existing state-of-the-art techniques for network traffic forecasting in both forecasting speed and accuracy.
\end{itemize}

\section{Related Works}
Existing works on network traffic forecasting mainly include traditional statistical methods and deep
learning (DL)-based approaches.
\subsection{Statistical Methods}
Statistical approaches predominantly utilize linear models for time series forecasting tasks. A widely adopted example is the ARIMA model \cite{moayedi2008arima}, which has been extensively applied in network traffic prediction. Xu et al. \cite{xu2014network} developed a prediction framework based on the Auto-Regressive Moving Average (ARMA) model, which integrates third-party monitoring systems to enhance prediction efficiency while minimizing network resource consumption. Despite their practicality, these linear models are often insufficient for capturing the intricate and nonlinear dynamics of real-world traffic patterns \cite{aouedi2025deep}. Furthermore, their reliance on complete historical datasets renders them ineffective in scenarios involving missing or incomplete data.

\subsection{Deep Learning Methods}
With the rapid advancement of artificial intelligence, DL has garnered significant attention in the field of network traffic forecasting. Numerous DL-based models have been introduced to achieve efficient and accurate network traffic forecasting\cite{alawe2018improving,vinayakumar2017applying,he2020meta}. Andreoletti et al. \cite{andreoletti2019network} applied a Diffusion Convolutional Recurrent Neural Network (DCRNN) to predict link-level traffic loads within a real-world backbone network. Xie et al. \cite{xie2024m} introduced a multi-range, multi-level spatio-temporal learning approach comprising three dedicated aggregation modules, each tailored to capture distinct temporal patterns (recent, daily, and weekly).  Despite their promising performance, these DL models are generally built on the assumption of fully available historical traffic data. This reliance on complete datasets significantly limits their applicability in practical scenarios, where network traffic data is normally sparse or incomplete.

Different from existing approaches to network traffic forecasting, our proposed GMF scheme employs a low-rank tensor framework \cite{zhang2016exact,deng2025graph} to achieve accurate forecasting in the presence of missing data. At the core of GMF is a pre-trained generative model that enforces the low-rank constraint required for TC. This design eliminates the need for computationally intensive tensor decompositions, thereby supporting real-time forecasting.

\section{Preliminaries and Problem Formulation}
We begin by defining the notations and preliminary concepts, and then proceed to describe the system model and problem formulation.
\subsection{Preliminaries}
\subsubsection{Notations}A tensor is a multi-dimensional generalization of matrices and vectors, typically represented as a multi-way array. Throughout this work, we denote tensors using calligraphic letters. For instance, a third-order tensor is represented as $\mathcal{T} \in \mathbb{R}^{n_1 \times n_2\times n_3}$. We use $\mathcal{T}(i,:,:)$, $\mathcal{T}(:,j,:)$, $\mathcal{T}(:,:,k)$ to denote the $i$-th horizontal, $j$-th lateral and $k$-th frontal slice of the tensor $\mathcal{T}$, and $\mathcal{T}(i,j,k)$ to denote its $(i,j,k)$-th entry, respectively. For convenience, we can define  $\mathbf{T}^{(k)} = \mathcal{T}(:,:,k)$. The Frobenius norm of a tensor $\mathcal{T}$ is defined as $||\mathcal{T}||_F=\sqrt{\sum_{i,j,k} |\mathcal{T}(i,j,k)|^2}$. $[n]$ represents the set of $[1,2,\cdots,n]$. We also denote $\widetilde{\mathcal T}$ as a tensor by performing the fast Fourier transform along the third mode of tensor ${\mathcal T}$, i.e., $\widetilde{\mathcal T}(i,j,:)=\text{fft}(\mathcal T(i,j,:))$. It can be written as $\widetilde{\mathcal T}(i,j,:)=\text{fft}(\mathcal T, [~], 3)$ in MATLAB code. Then ${\mathcal T}$ can also be derived from $\widetilde{\mathcal T}$, i.e., ${\mathcal T}(i,j,:)=\text{ifft}(\widetilde{\mathcal T}, [~], 3)$.

\subsubsection{Tensor Operations}We first present the t-product operation, which serves as a generalization of matrix multiplication for computing the product between two third-order tensors \cite{lin2022robust,hu2020network,wang2021robust}.

\begin{definition}(t-product \cite{Kilmer}) The t-product between $\mathcal{A} \in \mathbb{R}^{n_1 \times n_2\times n_3}$ and $\mathcal{B} \in \mathbb{R}^{n_2 \times n_4\times n_3}$ is defined as $\mathcal{C}\in \mathbb R^{n_1 \times n_4 \times n_3}=\mathcal{A} *\mathcal{B}$, where $
\mathcal{C}(i,j,:)=\sum_{\ell=1}^{n_2}\mathcal{A}(i,\ell,:)\odot\mathcal{B}(\ell,j,:),
$ and $\odot$ represent the circular convolution.
\end{definition}
Note that the t-product operation is similar to the matrix multiplication where the element-wise multiplication operation is replaced by the circular convolution of two tubes \cite{deng2020}. The t-product can also be regarded as the matrix multiplication in the Fourier domain, i.e.,
$\mathcal{C}=\mathcal{A}*\mathcal{B}$ is equal to $\widetilde{\mathcal C}=\widetilde{\mathcal A}\triangle\widetilde{\mathcal B}$, where $\triangle$ is the frontal-slice-wise product \cite{liu2021tensors}.

\begin{definition}(Tensor transpose \cite{Kilmer}) The transpose of an $n_1\times n_2\times n_3$ tensor $\mathcal{T}$ is a tensor of size $n_2\times n_1\times n_3$, denoted as $\mathcal{T}^{\top}$, which is obtained by transposing every frontal slice of the tensor and then sorting the transposed frontal slices 2 through $n_3$ in reversed order, i.e., $\mathcal{T}^{\top}(:,:,1)=(\mathcal{T}(:,:,1))^H$ and $\mathcal{T}^{\top}(:,:,n_3+k-2)=(\mathcal{T}(:,:,k))^H$ for $2\leq k\leq n_3$.
\end{definition}

Now we introduce tensor singular value decomposition (t-SVD) and present its details in Alg. \ref{Alg:t_svd}.

\begin{definition} (t-SVD and tensor tubal-rank \cite{Kilmer}) The t-SVD of a third-order tensor $\mathcal{T} \in \mathbb{R}^{n_1 \times n_2\times n_3}$ is given by $\mathcal T=\mathcal U*\mathcal S*\mathcal V^\top$. $\mathcal U$ and $\mathcal V$ are orthogonal tensors with size ${n_1 \times n_1\times n_3}$ and ${n_2 \times n_2\times n_3}$. The tubal-rank of tensor $\mathcal{T}$ is defined as the number of non-zero singular tubes of $\mathcal S$, while $\mathcal S$ is an orthogonal tensor.
\end{definition}

\begin{algorithm}[t]
\caption{t-SVD for Third-Order Tensors \cite{Kilmer}}
\label{Alg:t_svd}
\footnotesize 
\setlength{\lineskip}{2pt} 
\KwIn{${{\mathcal T}} \in {{\mathbb{R}}^{{n_1} \times {n_2} \times {n_3}}}$.}
\KwOut{${{\mathcal U}}, {{\mathcal S}}, {{\mathcal V}}$.}
${\widetilde{\mathcal T}}{\mathrm{ = fft}}\left( {{{\mathcal T}},[~],3} \right)$.\;
\For {$k = 1:~{n_3}$}{
    $[{\mathbf{ U} }, {\mathbf{ S} }{{, }}{\mathbf{ V}}] = {{\text{svd}(}}{\mathbf{\widetilde{ T}}^{(k)}}{{)}}$, ${\mathbf{\widetilde { U}}^{(k)}} = \mathbf{{U}}, {\mathbf {\widetilde S}^{(k)}} = \mathbf{{S}}, {\mathbf{\widetilde { V}}^{(k)}} = \mathbf{{V}}$.\;
}
${{\mathcal U}} = {\mathrm{ifft}}({\widetilde {\mathcal U}},[~],3), {{\mathcal S}} = {\mathrm{ifft}}({\widetilde {\mathcal S}},[~],3), {{\mathcal V}} = {\mathrm{ifft}}({\widetilde {\mathcal V}},[~],3)$.
\end{algorithm}

\subsection{System Model}
Consider a network comprising $n$ nodes. The end-to-end network traffic volume at a given time interval can be represented by an $n \times n$ matrix. As the spatial traffic matrix evolves over time, the sequence of traffic matrices forms a third-order tensor, denoted by $\mathcal{T} \in \mathbb{R}^{n \times n \times n_3}$, where $n_3$ represents the number of time intervals. This traffic tensor captures three dimensions: source nodes, destination nodes, and temporal intervals, as illustrated in Fig. \ref{fig:model}. Each entry $\mathcal T_{ijk}$ ($1\le i,j\le n,1\le k\le n_3$) corresponds to the traffic volume between origin node $i$ and destination $j$ during the $k$-th time interval.

\subsection{Problem Formulation}
Our objective is to achieve real-time and accurate network traffic forecasting in the presence of missing data. Specifically, consider a network consisting of $n$ nodes. Given a sequence of historical traffic matrices with missing entries, the task is to accurately predict future traffic matrices in real-time. Let $T_h$ denote the length of the historical sequence (historical length) and $T_p$ represent the length of the forecasting horizon (forecasting length). By stacking these traffic matrices together along the temporal dimension, we construct a third-order tensor $\mathcal{T} \in \mathbb{R}^{n \times n \times n_3}$, where $n_3 = T_h + T_p$.

\begin{figure}[t]
    \centering
    \begin{minipage}[b]{0.45\textwidth}
        \centering
        \includegraphics[width=7cm]{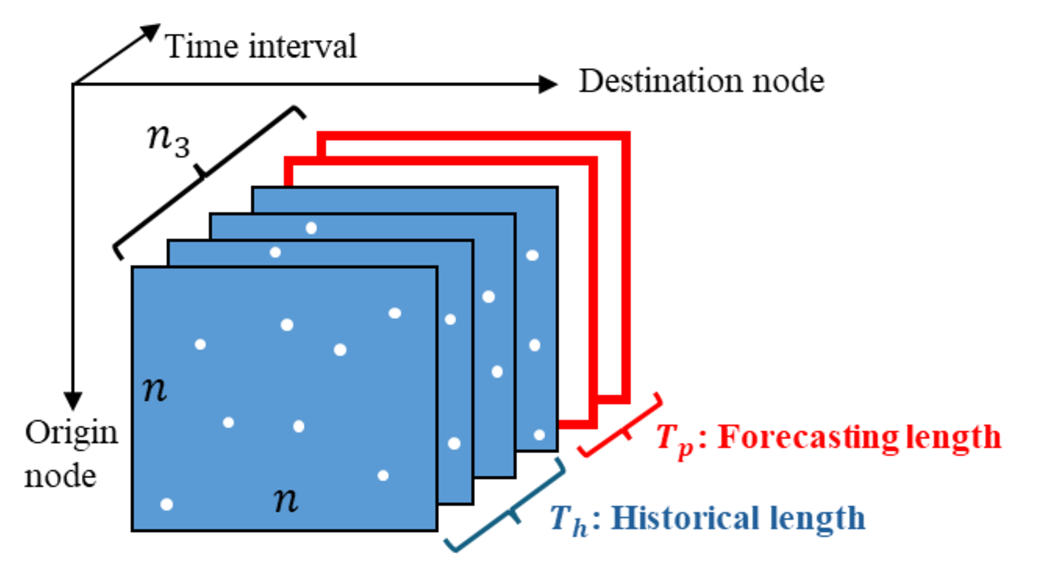}
        \caption{Tensor model for network traffic data. }
        \label{fig:model}
    \end{minipage}
    \hfill
    \begin{minipage}[b]{0.45\textwidth}
        \centering
        \includegraphics[width=5cm]{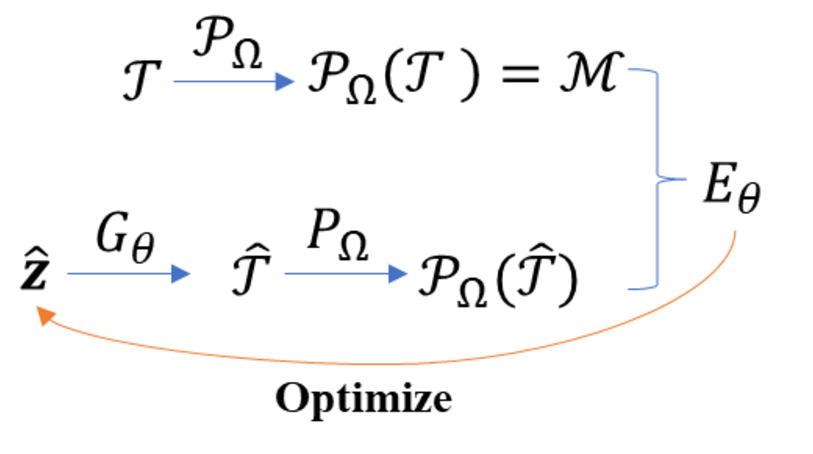}
        \caption{Illustration of the low-tubal-rank
TC framework based on genitive models.}
        \label{fig:illustration}
    \end{minipage}
\end{figure}

Accordingly, the forecasting task can be formulated as a tensor completion problem, where the goal is to predict the frontal slices $(\mathbf{T}^{(T_h+1)}, \mathbf{T}^{(T_h+2)}, \dots, \mathbf{T}^{(T_h+T_p)})$ in real-time based on incomplete historical observations $(\mathbf{T}^{(1)}, \mathbf{T}^{(2)}, \dots, \mathbf{T}^{(T_h)})$. An illustration of this process is provided in Fig.~\ref{fig:model}.

The low-rank TC problem is normally formulated as \cite{che2022fast,lu2019tensor}:
\begin{equation}\label{tnn}
\hat{\mathcal T}=\arg \min\limits_{\mathcal X\in \mathbb{R}^{n \times n \times n_3}} \| \mathcal M-\mathcal P_\Omega(\mathcal X)\|_F^2+\tau\|\mathcal X\|_\text{TNN},
\end{equation}
Note that $\|\mathcal X\|_\text{TNN}$ represents the tensor nuclear norm (TNN) of $\mathcal X$, which constrains that the recovered tensor is low-tuba l-rank. $\|\mathcal X\|_\text{TNN}$ is defined as the sum of the singular values of all the frontal slices of $\widetilde {\mathcal X}$, i.e., ${\mathrm{|}}{\mathrm{|}}{{\mathcal X}}{\mathrm{|}}{{\mathrm{|}}_{{\text{TNN}}}}{\mathrm{ = }}\sum\limits_{k = 1}^{{n_3}} {||\widetilde {{\mathcal X}}(:,:,k)||_*} $. $\hat{\mathcal T}$ is the recovered tensor, and $\mathcal M$ is measurement tensor (incomplete tensor) and $\tau>0$. We use $\Omega$ to denote the set of observed elements, $|\Omega|$ to denote the number of observed elements, and $\mathcal P_\Omega$ to denote the sampling operator, then we can define $\mathcal P_\Omega(\mathcal X)=\Phi\odot\mathcal  X$, where $\odot$ is the element-wise product (or Hadamard product), $\Phi$ is the indicator: $\Phi(i,j,k) = \begin{cases} 1, & \text{if } (i,j,k) \in \Omega, \\ 0, & \text{otherwise}. \end{cases}$ and we have $\mathcal M=\mathcal P_\Omega(\mathcal T)$.

The solution to \eqref{tnn} can be obtained through tensor singular value thresholding (t-SVT)\cite{che2022fast,lu2019tensor}, though this approach requires computationally expensive low-rank decompositions to compute TNN and suffers from a long runtime.

Recent work \cite{bora2017compressed} introduces compressed sensing using generative models (CSGM), which relaxes the traditional sparsity constraint in compressed sensing by leveraging a generative model as a structural prior. This approach motivates our key research question: \emph{Can a generative model achieve the tensor low-rank constraint in \eqref{tnn}, enabling real-time network traffic tensor completion}? To address this question, we have conducted empirical studies to evaluate the capability of generative models in producing low-tubal-rank tensor data (see Appendix \ref{sec:emp}). Now we explain our main idea.

\section{Generative Models for Low-Tubal-Rank TC}
\label{sec:frw}
We begin by introducing our main idea of low-tubal-rank TC framework using generative models. Suppose that a tubal-rank $r$ tensor $\mathcal{T} \in \mathbb{R}^{n_1 \times n_2 \times n_3}$ can be represented by a low-dimensional latent representation $\hat {\mathbf z}\in\mathbb R^{l}$, $l\ll {n_1  n_2  n_3}$. A pre-trained generative model (i.e., a generator) $G_\theta: \mathbb{R}^{l} \rightarrow \mathbb{R}^{n_1 \times n_2 \times n_3}$ is then used to map this latent representation $\hat{\mathbf{z}}$ to its corresponding tensor $\mathcal{T}$, such that:
\begin{equation}
\mathcal{T} =G_\theta(\hat{\mathbf z}).
\label{eq:generator}
\end{equation}

We now provide an intuitive explanation of how Equation~\eqref{eq:generator} operates. Let $\mathcal T=\mathcal U*\mathcal S*\mathcal V^\top$ be the t-SVD of $\mathcal{T}$ and define the singular vector $\mathbf v=[\mathcal{S}(1,1,:),\cdots,\mathcal{S}(r,r,:)]$. This vector $\mathbf{v}$ is constructed by stacking all the singular tubes from the diagonal of $\mathcal{S}$ into a single long vector.

It is known that these singular tubes encapsulate essential structural information of the tensor and have a significant influence on the performance of TC. Therefore, the singular vector $\mathbf v$ can be naturally interpreted as a latent representation of $\mathcal{T}$. By integrating the t-SVD formulation with \eqref{eq:generator}, we obtain:
\begin{equation}
\mathcal T=\mathcal U*\mathcal S*\mathcal V^\top=G_\theta(\mathbf v),
\label{eq:tsvd_generator}
\end{equation}
and $\hat{\mathbf z}\in\mathbb{R}^{l}=\mathbf v\in\mathbb{R}^{rn_3}$.  In this setting, $l=rn_3$ and the generator implicitly learns the orthogonal tensor components $\mathcal U$ and $\mathcal V$ during the pre-training phase, enabling it to map the singular vector $\mathbf v$ back into the original tensor space. During inference, the latent representation $\hat{\mathbf{z}}$ is optimized to approximate the ground truth singular vector $\mathbf{v}$. An illustration of this process is provided in Fig.~\ref{fig:illustration}.

The low-tubal-rank TC framework via generative models can be depicted by the following minimization function
\begin{equation}
\hat{\mathbf z}=\arg\min_{\mathbf z} E_\theta(\mathcal M,\mathbf z),
\label{eq:minE}
\end{equation}
where
\begin{equation}
E_\theta(\mathcal M,\mathbf z)=\|\mathcal M-\mathcal P_\Omega(G_\theta(\mathbf z))\|_F^2+\gamma\|G_\theta(\mathbf z)\|_\text{TNN}.
\label{eq:E_theta}
\end{equation}

\eqref{eq:E_theta} is aligned with \eqref{tnn}. Given that TNN is defined as the sum of singular values in the spectral domain, and $\mathbf{z}$ represents the potential singular vector of $G_\theta(\mathbf{z})$, we can approximate $\|G_\theta(\mathbf{z})\|_\text{TNN}$ using $\|\widetilde{\mathbf{z}}\|_1$ in practice, where $\widetilde{\mathbf{z}}$ is the spectral version of ${\mathbf{z}}$, defined by $\widetilde{\mathbf{z}}=[\text{fft}(\mathbf z(1,\cdots,n_3)),\cdots, \text{fft}(\mathbf z((r-1)n_3+1,\cdots,rn_3))]$. The parameter $\gamma$ serves to regulate the influence of the low-rank regularization.



\eqref{eq:minE} can be addressed using a naive gradient descent approach, starting from an initial point $ {\mathbf z}_1\sim p_{\mathbf z}(\mathbf z)$:
\begin{equation}
 {\mathbf z}_{t+1}\gets {\mathbf z}_{t}-\rho\frac{\partial E_\theta(\mathcal M,\mathbf z)}{\partial \mathbf z}\bigg|_{{\mathbf z}= {\mathbf z}_{t}},
\label{eq:v}
\end{equation}
where $\rho$ is the learning rate.

Unlike the conventional approach that directly optimizes the tensor $\mathcal T$ in the low-rank completion problem \eqref{tnn}, the optimization in \eqref{eq:minE} is carried out in the latent space of the variable $\mathbf z$. We also provide relevant theoretical analysis to reveal the error bound of the proposed framework in Appendix \ref{sec:theo}.

\section{Generative Models for Real-Time Network Traffic Forecasting with Missing Data}
We now elaborate the technical details of our GMF scheme for real-time network traffic forecasting with missing data, which comprises two key phases: a pre-training phase and an online inference phase.

\begin{figure}[t]
    \centering
    \includegraphics[width=12cm]{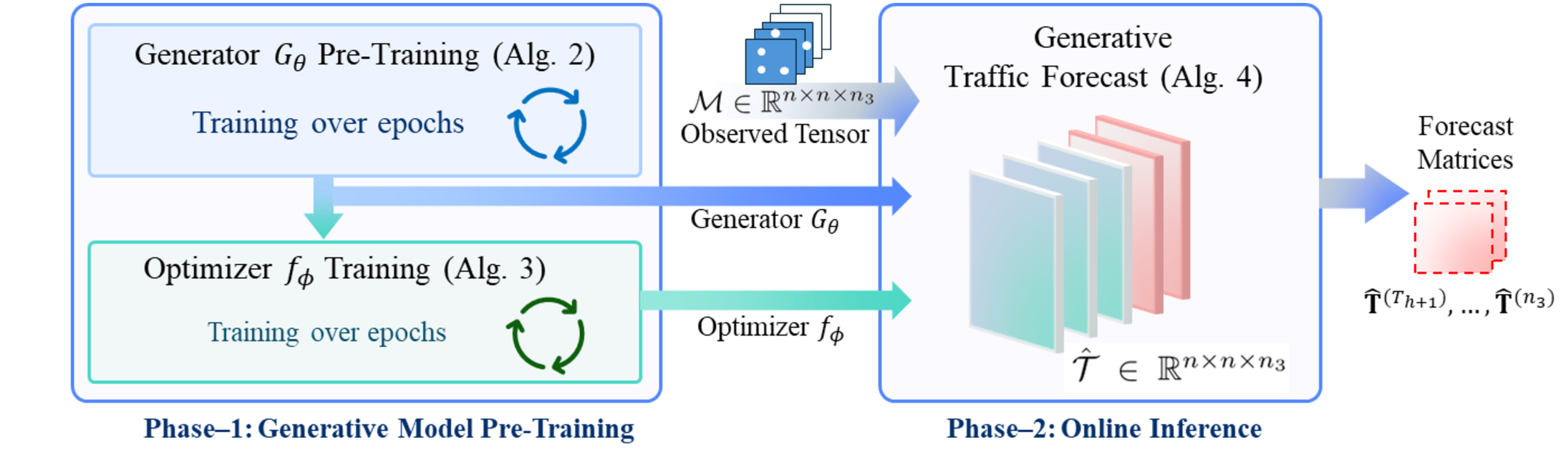}
    \caption{The framework of the GMF scheme}
\label{fig:gmf}
\end{figure}

\subsection{Phase-1: Pre-training of the Generative Model}
As presented in Section \ref{sec:frw}, a pre-trained generative model $G_\theta$ is required for TC and network traffic forecasting. The quality of $G_\theta$ highly impacts the forecasting accuracy. Here we first introduce the design and training of the generator $G_\theta$, and then we present a DL-based update rule to speed up the optimization of $\mathbf z$.

\subsubsection{Architecture Design for the Generative Model}
The architecture of the generator $G_\theta$ requires deliberate design to capture the low-tubal-rank structures of network traffic tensors. According to \eqref{eq:tsvd_generator}, we can observe that $G_\theta$ needs to learn the orthogonal tensor components $\mathcal U$ and $\mathcal V$ with t-product operations. In this setting, tensor layers (TL) \cite{newman2018stable,liu2024spectral} become natural choices. TL is a neural network that developed based on t-product. Its forward propagation is defined by
$\mathcal A_{j+1}=\sigma(\mathcal W_{j}*\mathcal A_{j}+\mathcal B_{j}),~\forall j=1,\cdots,N-1$, where $\sigma$ is an element-wise activation function, $N$ is the number of layers in the network, $\mathcal A_{j}\in\mathbb R^{\ell_{j}\times b\times a}$ is a feature tensor, $\mathcal W_{j}\in\mathbb R^{\ell_{j+1}\times\ell_j\times a}$ is a trainable weight tensor and $\mathcal B_{j}\in\mathbb R^{\ell_{j+1}\times b\times a}$ is a bias tensor.

In a TL, $\ell_j$ implicitly acts as the tensor tubal-rank. By setting $\ell_j$ to an appropriate small value, the output of a tensor layer is a low-tubal-rank tensor. Meanwhile, the t-product operation preserves and exploits multidimensional correlations within the data. This is in contrast to standard neural networks, which typically flatten tensors into vectors or matrices, destroying the multi-mode structure and losing access to low-tubal-rank information.

Therefore, we structure the generator $G_\theta$ with tensor layers
\begin{equation}
G_\theta(\cdot)=\sigma (\text{TL}(\sigma(\text{FC}(\cdot)))),
\end{equation}
where FC($\cdot$) represents the fully connected layer. We will show that this structure is more powerful than conventional neural networks like pure FC layers in the experiments.

\subsubsection{Training for the Generative Model}
We now describe the pre-training procedure for the generator $G_\theta$. The goal is to guide the generator to map $\mathbf z$ into a low-tubal-rank tensor space that approximates the actual network traffic tensor. Since the latent variable $\mathbf z$ to closely approximate the true singular vector $\mathbf v$ (as depicted in \eqref{eq:tsvd_generator}), a straightforward strategy for training $G_\theta$ is to use singular vectors as inputs and the corresponding complete traffic tensors as label, as outlined in Alg. \ref{Alg:train1}. The training loss in Alg. \ref{Alg:train1} consists of two components: (i) the first term strengthens the model's capacity to accurately translate singular vectors into the tensor domain; and (ii) the second term enforces a low-tubal-rank structure on the generated tensors by minimizing TNN. While the training routine in Alg. \ref{Alg:train1} does not employ adversarial learning, it well captures the mapping from singular vectors to the tensor domain.

\subsubsection{Latent Optimization Acceleration by DL}
Given a well-trained generator $G_\theta$, we can solve \eqref{eq:minE} by gradient descent  \eqref{eq:v} for TC and network traffic forecasting. However, standard gradient-based optimization typically requires executing a large number of iterations - often ranging from hundreds to thousands - along with multiple random restarts \cite{bora2017compressed,bojanowski2018optimizing} to obtain a sufficiently good $\hat{\mathbf z}$ .

Recent studies  \cite{andrychowicz2016learning,deng2025real} have demonstrated that the gradient descent process can be facilitated using DL. Motivated by this, we further adopt a DL-based optimizer $f_\phi$, parameterized by $\phi$, to replace the conventional update rule in \eqref{eq:v}. This substitution avoids multiple iterations and facilitates the optimization. The update of ${\mathbf z}$ is then performed as follows:
\begin{equation}
 {\mathbf z}_{t+1}=f_\phi(\mathcal M,\mathbf z_t),
\label{eq:optz}
\end{equation}
where $f_\phi(\mathcal M,\mathbf z_t)=\text{$\sigma$(FC($\sigma$(FC($\sigma$(FC(Cat(vec($\mathcal M), \mathbf z_t$)))))))}$ is a lightweight neural network. Here, vec$(\cdot)$ is vectorization operation and Cat$(\cdot)$ denotes the vector concatenation operation.

The following loss function is adopted to train $f_\phi$:
\begin{equation}\label{eq:lossf}
\text{loss}_f=\alpha\|\hat{\mathbf{z}}-\mathbf v\|_F^2+\beta\|G_\theta({\hat{\mathbf{z}}})-\mathcal T\|_F^2,
\end{equation}
where $\alpha$ and $\beta$ are weighting hyperparameters.  The first term encourages the optimized latent vector $\hat{\mathbf{z}}$ to approximate the true singular value vector $\mathbf{v}$, while the second term ensures that the generator $G_\theta$ reconstructs the target traffic tensor $\mathcal{T}$ accurately. The overall training procedure for $f_\phi$ is outlined in Alg. \ref{Alg:opt}.

\begin{figure}[t]
    \centering
    \begin{minipage}[t]{0.48\textwidth} 
        \begin{algorithm}[H] 
            \caption{Pre-Training of $G_\theta$}
            \label{Alg:train1}
            \footnotesize
            \setlength{\lineskip}{2pt}
            \KwIn{Complete network traffic tensor $\mathcal{T}\in\mathbb{R}^{n\times n\times n_3}$, max epoch $MaxEpoch$, parameter $\gamma_0$}
            \KwOut{Trained parameters $\theta$ of $G_\theta$}
            \textbf{Initialize}: Parameters $\theta$, $ep=1$\;
            \While{NOT converged \textbf{or} $ep\leq MaxEpoch$}{
                [$\mathcal{U}$, $\mathcal{S}$, $\mathcal{V}$] = t-SVD($\mathcal{T}$)\;
                $\mathbf{v} = [\mathcal{S}(1,1,:),\cdots,\mathcal{S}(n,n,:)]$\;
                loss = $\|G_\theta(\mathbf{v})-\mathcal{T}\|_F^2 + \gamma_0\|G_\theta(\mathbf{v})\|_\text{TNN}$\;
                Update $\theta$ to minimize loss, $ep=ep+1$\;
            }
            \Return $\theta$
        \end{algorithm}
    \end{minipage}
    \begin{minipage}[t]{0.48\textwidth}
        \begin{algorithm}[H]
            \caption{Training of $f_\phi$}
            \label{Alg:opt}
            \footnotesize
            \setlength{\lineskip}{2pt}
            \KwIn{Complete tensor $\mathcal{T}$, measurements $\mathcal{M}=\mathcal{P}_\Omega(\mathcal{T})$, latent $\mathbf{z}$, $MaxEpoch$, steps $K$, generator $G_\theta$}
            \KwOut{Optimizer parameters $\phi$}
            \textbf{Initialize}: $\phi$, $\mathbf{z}_1\sim\mathcal{N}(0,1)$, $\rho=0.01$, $ep=1$\;
            \While{NOT converged \textbf{or} $ep\leq MaxEpoch$}{
                \textbf{for} {$t=1:K$}  \textbf{do} {Update $\mathbf{z}_t$ by \eqref{eq:optz}}\;
                $\hat{\mathbf{z}} = \mathbf{z}_{K+1}$\;
                Compute singular vector $\mathbf{v}$ of $\mathcal{T}$ via t-SVD\;
                Update $\phi$ to minimize loss in \eqref{eq:lossf}, $ep=ep+1$\;
            }
            \Return $\phi$
        \end{algorithm}
    \end{minipage}
\end{figure}

\subsection{Phase-2: Online Inference}
With the pre-trained generative model $G_\theta$ and the optimizer $f_\phi$ in place, GMF can be deployed for real-time network traffic forecasting. The detailed computational procedure of GMF is outlined in Alg.~\ref{Alg:gmc}. In summary: (i) GMF leverages generative models to capture the low-tubal-rank structures inherent in network traffic data. By directly mapping a latent representation to a complete network traffic tensor, GMF can achieve real-time forecasting. (ii) GMF adopts a tensor-layer-based architecture, enabling efficient learning for low-tubal-rank structures. (iii) To accelerate inference, GMF replaces standard gradient descent with a DL-based update mechanism $f_\phi$.
\begin{algorithm}[t]
\caption{Online Inference: GMF for Real-Time Forecasting with Missing Data}
\label{Alg:gmc}
\footnotesize
\setlength{\lineskip}{2pt}
\KwIn{Observed tensor $\mathcal{M}\in\mathbb{R}^{n\times n\times n_3}$, latent $\mathbf{z}\in\mathbb{R}^{nn_3}$, steps $K$, generator $G_\theta$, optimizer $f_\phi$}
\KwOut{Forecast matrices $\hat{\mathbf{X}}$}
\textbf{Initialize}: $\mathbf{z}_1\sim\mathcal{N}(0,1)$\;
\textbf{for} {$t=1:K$} \textbf{do} 
    $\mathbf{z}_{t+1} = f_\phi(\mathcal{M},\mathbf{z}_t)$ \tcp*{Latent update via \eqref{eq:optz}}
$\hat{\mathcal{T}} = G_\theta(\mathbf{z}_{K+1})$\;
\Return $\hat{\mathbf{T}}^{(T_h+1)},\dots,\hat{\mathbf{T}}^{(n_3)}$
\end{algorithm}

\textbf{Remark: Fast inference for larger networks.} GMF can provide faster inference compared with the vanilla method, which directly solves \eqref{tnn} by t-SVT \cite{che2022fast,lu2019tensor}. Specifically, given an incomplete network traffic tensor with size $n\times n\times n_3$, t-SVT requires computation cost $\mathcal O(n^2n_3+n^3n_3)$ for TC \cite{lu2019low}, while our GMF scheme in Alg. \ref{Alg:gmc} requires $\mathcal O(n^2n_3+nn^2_3+n^2n_3)$. Our approach is more scalable for larger networks (i.e., with a larger number of nodes $n$).

\section{Performance Evaluation}
We test the performance of the proposed method through
extensive simulations.

\textbf{Datasets and preprocessing}: We perform experiments to evaluate the efficacy of the proposed GMF in network traffic forecasting using two real-world IP network traffic datasets: the Abilene dataset \cite{mekaouil2013} and the G$\acute{\text{E}}$ANT dataset \cite{uhlig2006providing}. The Abilene dataset captures end-to-end traffic patterns among 12 network nodes, recorded at 5-minute intervals over 168 days, resulting in 48,384 traffic matrices. An anomalously large value in the Abilene dataset (identified as an outlier) has been replaced with zero to ensure data consistency. The G$\acute{\text{E}}$ANT dataset contains traffic measurements across 23 nodes, collected every 15 minutes over 120 days, resulting in 10,772 traffic matrices.

For each dataset, the initial 80\% of consecutive traffic matrices are used for training, while the remaining 20\% are reserved for testing. We configure the forecasting task with a historical sequence length of $T_h = 10$ and a forecasting horizon of $T_p = 1$, resulting in a tensor length of $n_3 = T_h + T_p = 11$. By applying a sliding window of size 11, we generate multiple third-order tensors $\mathcal{T} \in \mathbb{R}^{n \times n \times 11}$ for both training and evaluation, where $n = 12$ for the Abilene dataset and $n = 23$ for the G$\acute{\text{E}}$ANT dataset. To simulate real-world scenarios with missing data, we randomly mask elements in the historical portion of each tensor (i.e., the first 10 frontal slices of size $n \times n \times 10$), with missing rates ranging from 0.1 to 0.9. For each generated tensor, we evaluate both the forecasting accuracy and computation time of the final traffic matrix $\mathbf{T}^{(n_3)}$.

\textbf{Baselines}: The following forecasting methods are used as baselines for performance comparison. Multi-range Multi-level Spatial-Temporal Learning (M$^2$STL) \cite{xie2024m}, Uncertainty-Aware Inductive Graph Neural Network (UIGNN) \cite{mei2023uncertainty}, Transformer \cite{vaswani2017attention}, and LSTM \cite{hochreiter1997long}. Since M$^2$STL, Transformer, and LSTM require complete historical data for effective forecasting, we first apply the TNN-ADMM tensor completion algorithm \cite{zhang2016exact} to impute missing values, followed by forecasting. To evaluate forecasting performance, we consider the following standard metrics: Mean Absolute Error (MAE) and Normalized Root Mean Squared Error (NRMSE), which are widely adopted in recent work \cite{xie2024m}. All neural network models are implemented using the PyTorch framework, and experiments are conducted on a computer equipped with an NVIDIA GeForce RTX 4090 GPU.

\begin{table}[t]
\caption{Forecasting Performance Comparison (MAE and NRMSE)}
\label{Tab:combined}
\centering
\setlength{\tabcolsep}{2pt} 
\begin{tabular}{l@{}|@{}ccccc@{}|@{}ccccc@{}} 
\hline
\hline
\noalign{\smallskip}
\multirow{3}{*}{Method} & \multicolumn{10}{c}{Missing Rate} \\
\cline{2-11}
 & \multicolumn{5}{c|}{Abilene} & \multicolumn{5}{c}{G$\acute{\text{E}}$ANT} \\
\cline{2-11}
 & 0.1 & 0.3 & 0.5 & 0.7 & 0.9 & 0.1 & 0.3 & 0.5 & 0.7 & 0.9 \\
\noalign{\smallskip}\hline\noalign{\smallskip}

\multicolumn{11}{l}{\textbf{MAE}} \\
\noalign{\smallskip}\hline\noalign{\smallskip}
GMF$_\text{TL}$ (Ours)& 0.0014 & \textbf{0.0015} & \textbf{0.0015} & \textbf{0.0016} & \textbf{0.0016} & \textbf{0.0015} & \textbf{0.0015} & \textbf{0.0015} & \textbf{0.0016} & \textbf{0.0016} \\
GMF$_\text{FC}$ & 0.0020 & 0.0020 & 0.0021 & 0.0021 & 0.0022 & 0.0113 & 0.0116 & 0.0122 & 0.0125 & 0.0128 \\
M$^2$STL & 0.0015 & 0.0018 & 0.0019 & 0.0023 & 0.0026 & 0.0016 & 0.0017 & 0.0018 & 0.0019 & 0.0020 \\
UIGNN & 0.0017 & 0.0020 & 0.0023 & 0.0028 & 0.0031 & 0.0065 & 0.0070 & 0.0077 & 0.0089 & 0.0134 \\
LSTM & \textbf{0.0013} & \textbf{0.0015} & 0.0016 & \textbf{0.0016} & 0.0017 & 0.0024 & 0.0024 & 0.0026 & 0.0027 & 0.0029 \\
Transformer & 0.0022 & 0.0023 & 0.0025 & 0.0027 & 0.0028 & 0.0039 & 0.0040 & 0.0043 & 0.0044 & 0.0046 \\
\noalign{\smallskip}\hline\noalign{\smallskip}

\multicolumn{11}{l}{\textbf{NRMSE}} \\
\noalign{\smallskip}\hline\noalign{\smallskip}
GMF$_\text{TL}$ (Ours)& \textbf{0.0122} & \textbf{0.0123} & \textbf{0.0124} & \textbf{0.0128} & 0.0135 & \textbf{0.0045} & \textbf{0.0045} & \textbf{0.0046} & \textbf{0.0047} & \textbf{0.0047} \\
GMF$_\text{FC}$ & 0.0129 & 0.0132 & 0.0133 & 0.0136 & 0.0137 & 0.0051 & 0.0052 & 0.0052 & 0.0053 & 0.0054 \\
M$^2$STL & 0.0516 & 0.0529 & 0.0553 & 0.0632 & 0.0783 & 0.0781 & 0.0782 & 0.0783 & 0.0787 & 0.0792 \\
UIGNN & 0.0629 & 0.0713 & 0.0802 & 0.0933 & 0.1130 & 0.1209 & 0.1752 & 0.2221 & 0.2469 & 0.2732 \\
LSTM & 0.0125 & 0.0127 & 0.0128 & 0.0130 & \textbf{0.0133} & 0.0046 & 0.0049 & 0.0050 & 0.0053 & 0.0056 \\
Transformer & 0.0135 & 0.0136 & 0.0136 & 0.0137 & 0.0138 & 0.0054 & 0.0056 & 0.0057 & 0.0057 & 0.0059 \\
\hline
\noalign{\smallskip}\hline
\end{tabular}
\end{table}

\subsection{Ablation Studies and Forecasting Accuracy}
We begin by evaluating the effectiveness of the tensor-layer architecture used in the generator. To this end, we define a variant of GMF composed solely of FC layers, expressed as $G_\theta(\cdot)=\delta\text{(FC($\delta$(FC($\cdot$))))}$, and refer to it as GMF$_\text{FC}$. The original GMF model utilizing tensor layers is denoted as GMF$_\text{TL}$. We compare the forecasting accuracy of GMF$_\text{TL}$ and GMF$_\text{FC}$ to assess the impact of the tensor-based design.

Table \ref{Tab:combined} reports the forecasting accuracy (MAE and NRMSE) across different data missing rates in historical data. It can be observed that GMF$_\text{TL}$ consistently outperforms GMF$_\text{FC}$, which relies solely on FC layers. This performance gap highlights the advantage of incorporating tensor layers in effectively capturing the low-rank structure inherent in network traffic data.

Comparing forecasting errors in Table~\ref{Tab:combined}, we observe: 
(i) GMF consistently outperforms competitors across most missing rates. 
(ii) On Abilene, LSTM occasionally matches or exceeds GMF, benefiting from TNN-ADMM's effective imputation of complete input data. However, TNN-ADMM's weaker recovery on G$\acute{\text{E}}$ANT limits LSTM's performance there. 
(iii) UIGNN, designed for transportation traffic, shows suboptimal performance without domain adaptation. 
Overall, GMF demonstrates superior robustness for network traffic forecasting.

\subsection{Real-time Performance}
We evaluate the real-time performance of different methods in Table~\ref{Tab:time}. UIGNN, GMF$_\text{TL}$, and GMF$_\text{FC}$ achieve <100 ms inference times, enabling real-time use. However, UIGNN and GMF$_\text{FC}$ sacrifice too much accuracy for speed. Traditional methods (M$^2$STL, LSTM, Transformer) require additional imputation steps (e.g., TNN-ADMM), increasing computation time significantly. GMF provides the optimal balance between speed and accuracy for real-time traffic forecasting.

\begin{table}[t]
\caption{Computational Time Comparison.}
\label{Tab:time}
\centering
\begin{tabular}{l|cccccc}
\hline
\hline
Dataset  & \cite{zhang2016exact}+M$^2$STL & \cite{zhang2016exact}+Transformer & \cite{zhang2016exact}+LSTM & UIGNN & GMF$_\text{FC}$ & GMF$_\text{TL}$\\
\hline
Abilene & $>2$ s & $>1$ s & $>1$ s & $<100$ ms & $<100$ ms & $<100$ ms \\
G$\acute{\text{E}}$ANT & $>5$ s & $>3$ s & $>3$ s & $<100$ ms & $<100$ ms & $<100$ ms \\
\hline
\hline
\end{tabular}
\end{table}

\section{Conclusion}
In this work, we proposed a Generative Model approach for real-time network traffic forecasting with missing data (GMF). GMF formulates the forecasting problem as a tensor completion task and leverages a generative model as a structural surrogate for conventional tensor low-rank assumptions. The model enables the generation of complete network traffic tensors from compact latent representations - thereby streamlining the forecasting process. Additionally, we adopted a tensor-layer architecture in the generative model, which better captures the low-tubal-rank structures in network traffic. Extensive evaluations on real-world network traffic datasets demonstrate that GMF reduces estimation time while preserving high accuracy.


\bibliographystyle{ACM-Reference-Format}
\bibliography{references}

\appendix

\section{Empirical Study}
\label{sec:emp}
The tensor low-rank properties of network traffic data have been widely discussed \cite{xie2018accurate,deng2023}. To verify if the generative models can capture the tensor low-rank constraint in network traffic tensors,
we investigate the low-tubal-rank properties of outputs produced by such models.

Generative artificial intelligence represents a transformative advancement in modern machine learning and has attracted widespread academic and industrial attention \cite{goodfellow2014generative,gpt4,deng2025real}. Recent investigations have shown the potential of generative models to effectively handle non-convex sparse vector structures in the context of compressed sensing \cite{bora2017compressed,wu2019deep}. Inspired by these developments, we examine whether generative models can also capture the non-convex low-rank structures that are commonly found in tensors.

We begin by evaluating the low-tubal-rank properties of tensors produced by well-established generative models. Specifically, we focus on two representative architectures: (i) Deep Convolutional Generative Adversarial Networks (DCGAN) \cite{radford2015unsupervised}, which take a Gaussian noise vector $\mathbf z \in\mathbb R^{120}$ and generate an image (tensor) of dimensions $64\times 64\times 3$. (ii) Progressive Growing of GAN (PGAN) \cite{karras2018progressive}, which takes a Gaussian noise vector $\mathbf z \in\mathbb R^{512}$ and generates an image (tensor) of dimensions $256\times 256 \times 3$.

\begin{figure}[t]
    \centering
    \includegraphics[width=9cm]{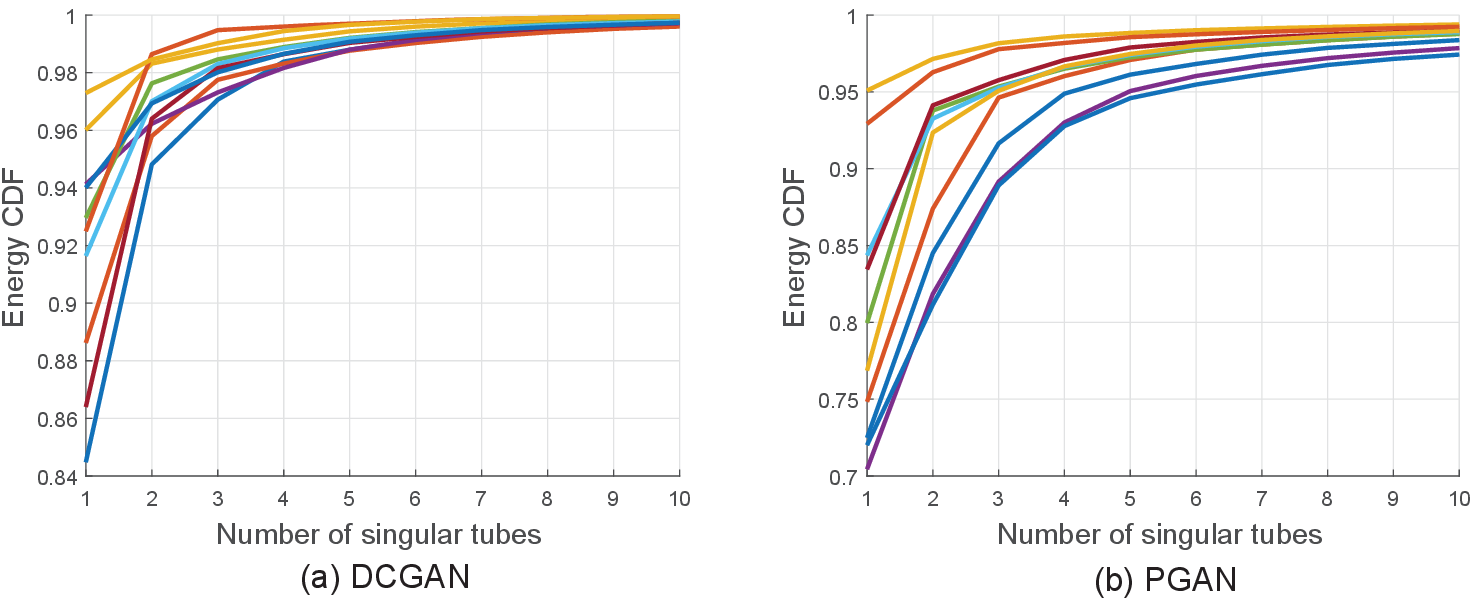}
    \caption{Energy CDFs of singular tube of tensors generated by generative models.}
    \label{fig:gan_energy}
\end{figure}

For each generative architecture, we generate input samples from a Gaussian distribution and use the output tensors for rank analysis. This experiment is repeated 10 times to obtain a collection of 10 tensors, from which we compute the t-SVD and the cumulative distribution functions (CDFs) of their singular tube energy. The energy of each singular tube in $\mathcal S$ bt t-SVD is defined as its $\ell_2$ norm. The results are presented in Fig.~\ref{fig:gan_energy}, which displays the energy distribution of singular tubes for both DCGAN- and PGAN-generated tensors. Remarkably, both models exhibit strong low-tubal-rank tendencies. For instance, in the case of DCGAN, the top two singular values alone account for over 94\% of the total energy.

These findings clearly demonstrate the inherent ability of generative models to represent and preserve tensor low-tubal-rank structures. Encouraged by this observation, we are motivated to apply such generative models to the task of low-tubal-rank TC and network traffic forecasting.

\section{Theoretical Analysis}
\label{sec:theo}
We present our main theoretical results for the generative model-based low-tubal-rank TC framework (i.e., Section \ref{sec:frw}), which are developed using compressive sensing techniques. Before proceeding, we introduce several essential definitions and notational conventions. Following the terminology in \cite{bora2017compressed,jalal2020robust}, for two functions $f_1(n)$ and $f_2(n)$, we write $f_1(n)=\Upsilon (f_2(n))$ to indicate the existence of a positive constant $c_3$ and a natural number $n_1$ such that for all $n\ge n_1$, the inequality $|f_1(n)|\ge c_3f_1(n)$ holds. We use $a\sim\mathcal N(\mu,\sigma^2)$ to denote that the variable $a$ satisfies a Gaussian distribution. For convenient discussion, we assume $n_1=n_2=n$. We also define $\mathbf a^l$ as the vectorization of the lateral slice of tensor $\mathcal A\in \mathbb{R}^{n \times n \times n_3}$ , i.e., $\mathbf a^l=[\mathcal A^{(1)}{(:,l)},\cdots,\mathcal A^{(n_3)}{(:,l)]^\top},~l\in[n]
$.

\begin{lemma}
\label{lem:sp}
Let $\mathcal{T} \in \mathbb{R}^{n \times n \times n_3}$ be a tensor with tubal-rank $r$. Then, there exists a matrix $\mathbf{A}$ such that the sampled tensor $\mathcal{M} = \mathcal{P}_\Omega(\mathcal{T})$ can be expressed in vectorized form as:
\[
\mathbf{m}^l = \mathbf{A} \mathbf{g}^l,\quad \forall~l \in [n],
\]
where each vector $\mathbf{g}^l$ is $rn_3$-sparse\footnote{A vector is said to be $s$-sparse if it contains at most $s$ nonzero entries.}.
\end{lemma}

The proof of Lemma \ref{lem:sp} can be derived by extending Theorem 2 in \cite{deng2023} to the t-product case.

\begin{definition}(Sampling Variance Condition)
Let $\mathcal{T} \in \mathbb{R}^{n \times n \times n_3}$ be a tensor, and let $\Omega$ denote the set of observed entries. The sampling operation is given by $\mathcal{M} = \mathcal{P}_\Omega(\mathcal{T})$, which can be expressed in vectorized form as: $\mathbf{m}^l = \mathbf{A} \mathbf{g}^l, \quad \forall~l \in [n]$,
for some matrix $\mathbf{A}$. If the entries of $\mathbf{A}$ are independently sampled from a Gaussian distribution, i.e., $\mathbf{A}(i,j) \sim \mathcal{N}(0, \sigma^2)$, then we say that the measurement tensor $\mathcal{M}$ satisfies the \textit{sampling variance condition} with parameter $\sigma^2$.
\label{Def:1}
\end{definition}

Given Definition \ref{Def:1}, we provide the following lemma:

\begin{lemma}
Let $\mathcal T\in\mathbb R^{n\times n\times n_3}$ be a rank-$r$ tensor. Suppose that $G_\theta:~\mathbb R^{rn_3}\to\mathbb R^{n\times n\times n_3}$ is a generative model from a $d$-layer neural network using ReLU activations, and we observe a subset of elements (denoted by $\Omega$) of $\mathcal T$ uniformly at random. Let $|\Omega|=\mathcal O(rn_3d\log(n^2n_3))$ and let the corresponding measurement tensor satisfy the variance condition with parameter $\frac{1}{|\Omega|}$. For any observations $\mathcal M=\mathcal P_\Omega(\mathcal T)$, there exists a $\gamma\ge0$ such that if $\hat {\mathbf z}$ minimizes \eqref{eq:E_theta} to within additive $\epsilon$ of the optimum, then with $1-e^{-\Upsilon({|\Omega|})}$ probability,
\begin{equation}
\|G_\theta(\hat {\mathbf z})-\mathcal T\|_F\le6\min\limits_{{\mathbf z}^*\in\mathbb R^{rn_3}}\|G_\theta({\mathbf z}^*)-\mathcal T\|_F+2\epsilon.
\end{equation}
\label{lem1}
\end{lemma}

We omit the proof of Lemma \ref{lem1} here, which can be derived by extending Theorem 1.1 in \cite{bora2017compressed} from the vector to the tensor space.

\end{document}